\documentclass[preprint,12pt]{elsarticle}

\usepackage{amssymb}
\usepackage{lineno}

\usepackage{multirow}
\usepackage[flushleft]{threeparttable}
\usepackage{booktabs}

\journal{Astroparticle Physics}

\begin{document}

\begin{frontmatter}



\title{Design, optimization and characterization of the light concentrators of
the single-mirror small size telescopes of the Cherenkov Telescope Array.}


\author[dpnc]{J. A. Aguilar}
\author[dpnc]{A. Basili} 
\author[dpnc]{V. Boccone}
\author[dpnc]{F. Cadoux}
\author[dpnc]{A. Christov}
\author[dpnc]{D. della Volpe}
\author[dpnc]{T. Montaruli}
\author[nca]{{\L}. P{\l}atos}
\author[dpnc]{M. Rameez} 
\author{for the SST-1M sub-Consortium.}
\address[dpnc]{D\'epartement de physique nucl\'eaire et corpusculaire Universit\'e de Gen\`eve, 1211 Geneva, Switzerland}
\address[nca]{Space Research Centre, Polish Academy of Sciences, ul. Bartycka 18A, 00-716 Warsaw, Poland}

\begin{abstract}

The focal-plane camera of $\gamma$-ray telescopes frequently uses light
concentrators in front of light sensors. The purpose of these concentrators is
to increase the effective area of the camera as well as to reduce the stray
light coming at large incident angles. 
These light concentrators are usually based on the
Winston cone design. In this contribution we present the design of an
hexagonal hollow light concentrator with a lateral profile optimized using a cubic
B\'ezier function to achieve a higher collection efficiency in the angular region
of interest. 
The design presented here is optimized for a Davies-Cotton telescope with
primary mirror of about 4 meters of diameter and focal length of 5.6 m. The
described concentrators are part of an innovative camera made up of
silicon-photomultipliers sensors, although a similar approach can be used for
other sizes of single-mirror telescopes with different camera sensors, including photomultipliers. The
challenge of our approach is to achieve a cost-effective design suitable for standard industrial productions of
both the plastic concentrator substrate and the reflective coating. At the same time we maximize
the optical performance. In this paper we also describe the optical set-up to measure the absolute collection efficiency of the
light guides and demonstrate our good understanding of the measured data using a
professional light tracing simulation.  

\end{abstract}


\end{frontmatter}


% \linenumbers


\section{Introduction}

The Cherenkov Telescope Array project (CTA)~\cite{CTA} will be the next generation observatory
in $\gamma$-ray astronomy. It will consist of two arrays of imaging atmospheric Cherenkov telescopes (IACTs) of
different sizes to be installed in the two hemispheres. The southern array will be composed of few
Large Size Telescopes (LSTs) of $\sim 24$~m diameter, about 25 Middle Size
Telescopes (MSTs) of about 12~m diameter and about 70 Small Size Telescopes
(SSTs) of $\sim 4$ m diameter. An arrangement over an area of the order of few
km$^2$ will make it possible to cover about 2 decades in energy, from about 20 GeV to 300
TeV, with an improved sensitivity by about a factor of 15 compared to existing
experiments which also operate in a smaller energy range.

Reflective light concentrators are a common element used in IACTs in
order to increase the collection area of camera pixels~\cite{HESS, MAGIC, VERITAS, FACT}. The cross-section
of these concentrators is usually hexagon-shaped so that they can be arranged in
an array with equal distance between all pixel centers. If their mutual distance is kept constant along the camera, the response of the pixels for different
orientations of the shower images is geometrically unbiased.
%% I do not think this is well explained nor the reference seems to exist: from This equal-distance pixel array is useful for the analysis point of view where image cleaning algorithms are used to discriminate the signature on the camera from $\gamma$-ray showers against proton induced showers~\cite{MAGIC}. 
These light concentrators are designed to maximize the collection efficiency of
incoming rays within the angular range subtended by the primary mirror and the
camera while at the same time limiting the amount of light from the night sky
background (NSB) at larger angles. The NSB arises from airglow, stars, nearby cities, etc. and
usually reaches the camera sensors directly without reflecting first on the telescope's primary mirror. These trajectories can be rejected geometrically by optimizing the shape of the light concentrator. The spectrum of the NSB is also different from the Cherenkov spectrum, the latter peaks at $\sim$~330 nm and begins to be dominated at wavelengths larger than 600 nm by the NSB that extends in the infrared. Therefore the wavelength dependency of the reflectivity material of the light concentrator can be optimized to further reduce this background.

The Small Size Telescopes of CTA are characterized by a wide field-of-view (FoV) of
about $9^{\circ}$ which is essential for the physics goals of these telescopes, such as the Galactic Plane survey, study of spectra to
discriminate hadronic against electromagnetic emissions, and the search for the Galactic PeVatrons which produce the Galactic cosmic rays up to the
{\it knee} at a few PeV~\cite{cta_science}. In the following we will explore the design of a light concentrator
suitable for a single mirror IACT and so we present it as a feasible solution for the SSTs. Particularly we consider the Davies-Cotton design which provides a good
imaging over a wide FoV. The dispersion in the arrival time of photons
introduced by this design for small dish telescopes is negligible compared to
the intrinsic dispersion of photons and therefore it is a suitable solution for
the SSTs. 
We pay particular attention to the
cost-effectiveness of our proposal and to the industrial producibility of the
light concentrators given the large number (around 70) of the SSTs planned for CTA.
Assuming the number of pixels for each camera to be around 1'300, the total
number of light concentrators to be produced for the SSTs in CTA will be of the
order of 91'000. Therefore it is clear that the design has not only to address the optical performance but has also 
the industrial producibility.

The outline of the paper is the following: in Section~\ref{sec:design} we review the Davies-Cotton telescope design and evaluate the parameters needed for a wide FoV telescope with a
$\sim$4~m diameter dish. We review the requirements of light
concentrators for this kind of design. In Section~\ref{sec:simulations} we show how, using a professional ray-tracing
simulation, we can optimize the geometrical shape of the light concentrator
using cubic B\'ezier curves. We compare this design with the Compound Parabolic Concentrator (CPC), also called Winston's Cone~\cite{winston}.
Even if solid concentrators can achieve larger compression factors than
hollow ones~\cite{FACT, FACT2, Bretz}, the former start to suffer from limited transmissivity
due to the thickness of $\sim$37~mm necessary in this design for the SST. In addition,
hollow cones allow us to use a reflective coating optimized for
high reflectivity in the wavelength region of interest. We consider a few
coating options and their impact on the collection efficiency.
Section~\ref{sec:measurements} shows the optical measurements performed with
the first prototypes and their comparison with simulations. Conclusions are
given in Sec.~\ref{sec:conclusions}.

\section{Design of a Davies-Cotton telescope.} 
\label{sec:design} 

As already stated, the telescope design has a primary mirror diameter,
{\it D} of 4 m, a FoV of 9$^\circ$, and a focal-length {\it f} of  5.6~m.
The focal-length is usually selected to keep the {\it f/D} in the range
1.2$-$1.4 where optical aberrations are small enough to facilitate good imaging
on the camera. % guaranteing a good $\gamma$-ray direction reconstructions. 

\begin{figure}[ht]
\hspace{-1cm}
  \includegraphics[ width=1.1\textwidth]{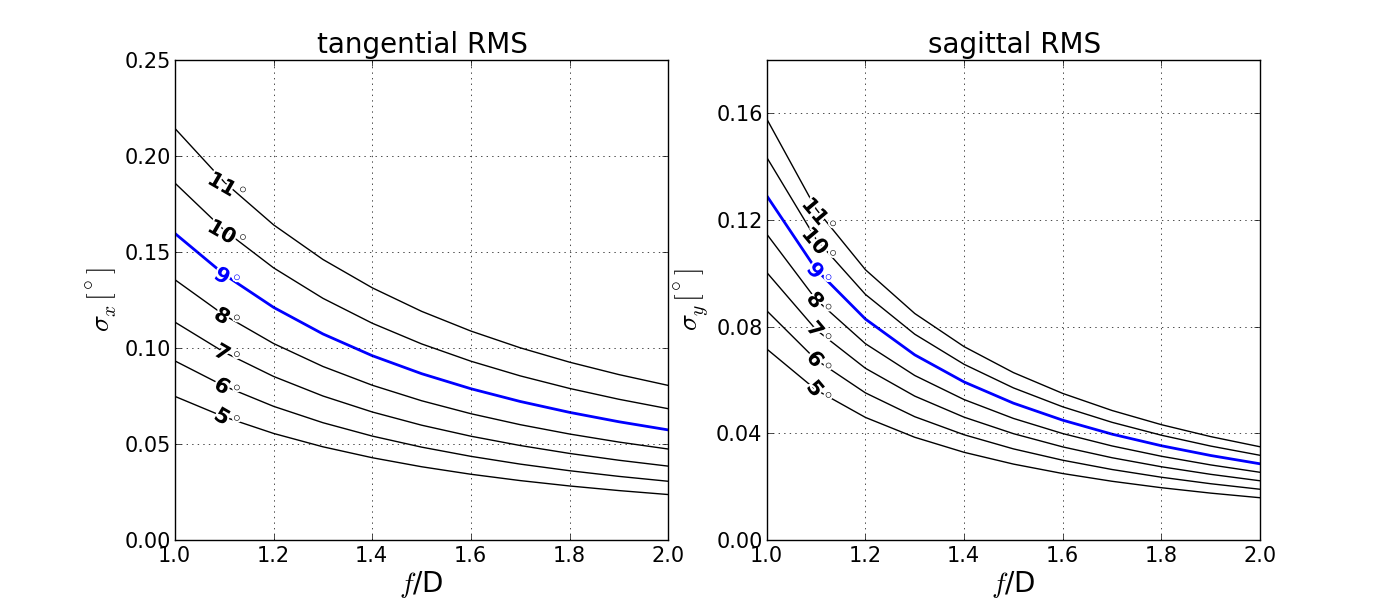}
\caption{RMS of the PSF for both the tangential and sagittal direction for a Davies-Cotton telescope as a function of $f/D$ for different FoV. Expressions are taken from~\cite{dc}. Indicated in blue is the line for a FoV of 9$^{\circ}$.}
\label{fig:rms}
\end{figure}

The IACT reflector is usually segmented into individual mirror facets and this
finite-facet-size enhances the optical aberrations of the system.
However, if one ignores these effects and considers only the global aberrations
of the Davies-Cotton design, the Point Spread Function (PSF) of photons can be
described analytically~\cite{dc}. The Root Mean Square (RMS) of these analytical
expressions for a Davies-Cotton design are shown in Fig.~\ref{fig:rms} for
both the tangential and sagittal direction as a function the {\it f/D} of the telescope.

%JA: To comment on a previous comment from Mimmo. These sagittal and tangential 
%directions are known directions in optical systems.

The RMS of the PSF can be used to define the optimum angular size of the pixel
of the camera. A compromise is to take an angular pixel size, {\it p}, that is 4
times the minimum of the two RMS. This guarantees the inclusion of 95.4\% of the
photons from a source assuming a gaussian distribution for the PSF. 

\begin{figure}[ht]
\centering
  \includegraphics[ width=1\textwidth]{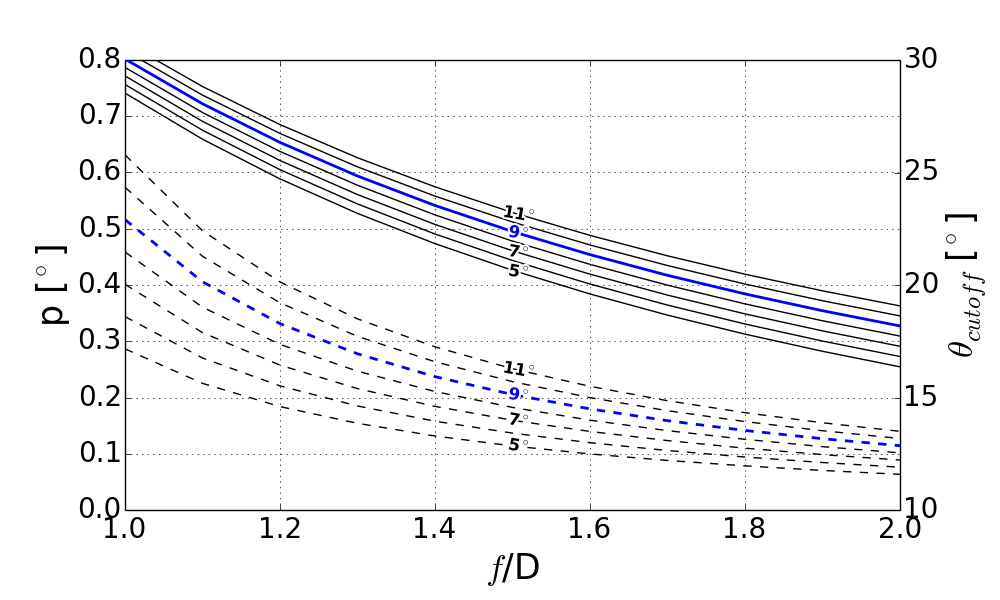}
\caption{We show on the left vertical scale and with dashed lines the angular pixel size, {\it p}, and on the right scale with solid lines the cutoff angle, $\theta_{cutoff}$, as a function of $f/D$ and for different FoV. Indicated in blue are the lines for FoV $ = 9^{\circ}$.}
\label{fig:pixelsize}
\end{figure}

As already mentioned, one of the goals of the light concentrators is to reduce the stray light from
the NSB. The camera is pointing to the dish mirror of the
telescope and therefore any ray of light that arrives at the camera without
undergoing any reflection on the mirror is considered to be stray light. The
light concentrators can veto the light that comes with an angle greater than the
one subtended by the mirror dish and the camera. This angle is usually referred to
as the cutoff angle $\theta_{cutoff}$. Both the angular pixel and the cutoff angle depend on $f/D$ and the FoV. In particular $\theta_{cutoff}$ is given by:

\begin{equation}
\theta = \arctan\left(\frac{D + D_{c}}{2f}\right)
\label{cutoff}
\end{equation}

\noindent where, {\it D}, refers to the mirror diameter and {\it D$_{c}$} to the
camera diameter. Figure~\ref{fig:pixelsize} shows both the angular pixel size,
{\it p} and the cutoff angle $\theta_{cutoff}$, as a function of the the {\it
f/D} for a telescope of 4 m-dish mirror. For our telescope, the ratio {\it f/D}
was selected to be 1.4.

\begin{figure}[ht]
\centering
  \includegraphics[ width=0.6\textwidth]{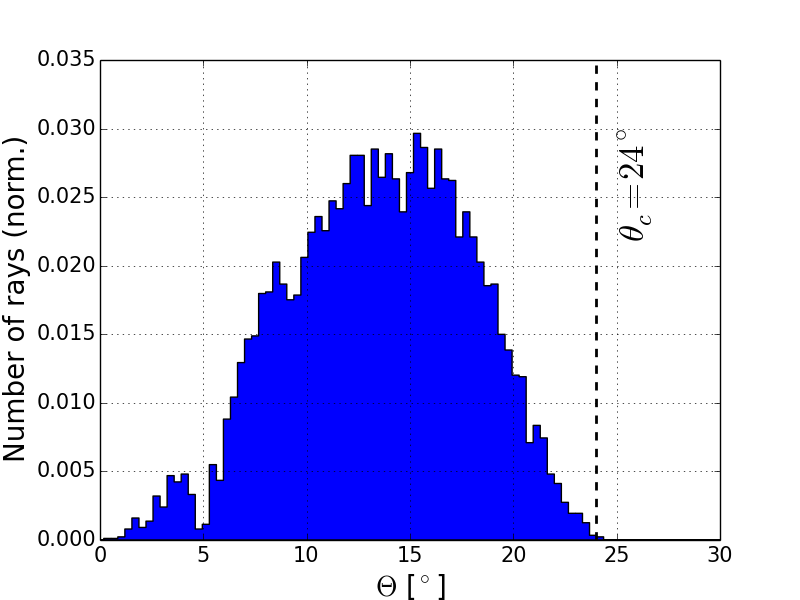}
\caption{Angular distribution on the camera plane of light reflected on the mirror. At $\sim$5$^{\circ}$ the effect of the camera mounting shadowing is visible. No other shadowing effects are considered in this plot.}
\label{fig:anglescamera}
\end{figure}

 As visible in Fig.~\ref{fig:pixelsize} the resulting angular pixel size is about
 0.25$^\circ$ and the resulting desirable cutoff angle for the light concentrator is $\sim
 24^{\circ}$.  In reality, cameras from IACTs might not have a circular shape. In order to verify that the geometrical estimated cutoff angle from Eq.~\ref{cutoff} covers all the incident angles of light reflecting on the mirror a raytracing simulation of the telescope optical system was performed. Figure~\ref{fig:anglescamera} shows the angular distribution of rays arriving on to the camera plane after reflecting on the primary mirror. As confirmed by the simulation, no ray of light reaches the camera plane at angles $> 24^{\circ}$.
 
%The angular pixel size and the cutoff angle constrain the geometrical dimensions of the
 The angular pixel size fixes the input aperture area of the concentrator, $a_{in}$,  while the cutoff angle fixes the exit aperture
 area $a_{out}$. In Ref.~\cite{winston}, R.~Winston proposed the two-dimensional
 Compound Parabolic Concentrator (CPC) as an ideal concentrator that reaches the maximum concentrator ratio $C_{max}$ given by:

\begin{equation}
C_{max} = \frac{a_{in}}{a_{out}} = \frac{1}{\sin^2\theta_{cutoff}}
\label{c}
\end{equation}
 
This ideal concentrator achieves a 100\% collection efficiency for incident
angles $\theta < \theta_{cutoff}$ while for angles $\theta > \theta_{cutoff}$ no
incident ray reaches the exit aperture of the concentrator. In three-dimensions
however the CPC design departs from the ideal case as rays that enter
tangentially to the side walls may follow spiral paths and may suffer many
reflection losses reducing the collection efficiency. 

The idea of using B\'ezier  curves in the profile shape of the concentrator in order to improve the collection efficiency was first demonstrated in~\cite{Akira}. Here we follow the same approach where starting from a CPC design the geometry is optimized using a cubic B\'ezier curve. The details of this optimization are described in the next section.

\section{Simulations} 
\label{sec:simulations} 
For the simulation and geometrical
optimization of the light concentrator we used the
Zemax\footnote{www.radiantzemax.com} software. This package is designed to
provide a tool for optical and illumination designs. It also includes an
optimization engine which makes it possible to reshape the sides of the light
concentrator to achieve the desirable figure of merit.

\subsection{Geometrical optimization}

\begin{figure}[ht]
\centering
  \includegraphics[ width=1\textwidth]{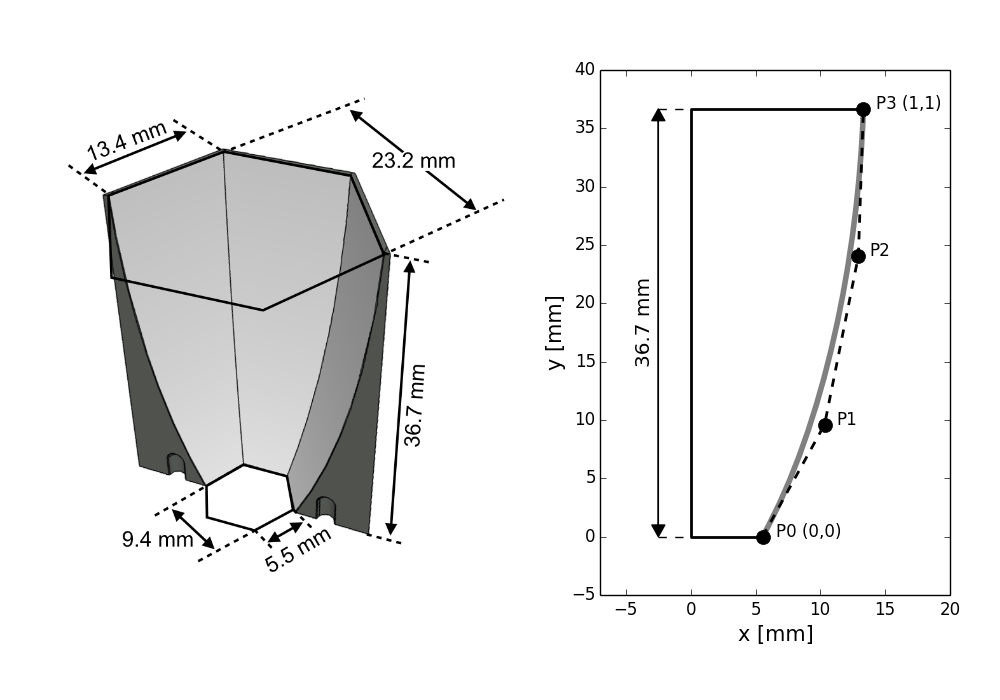}
\caption{CAD model of light concentrator with the dimensions indicated (left
panel) and cross section of the side of the concentrator (right panel).
Indicated are the four points of the cubic B\'ezier shape: Points $\vec{P}_0$
and $\vec{P}_3$ are fixed in order to keep the entrance and exit apertures
dimensions fixed. The points $\vec{P}_1$ and $\vec{P}_2$ are free parameters
during the optimization procedure.}
\label{fig:dimensions}
\end{figure}

As explained in Sec.~\ref{sec:design} we started from the CPC design which means
that the size of the entrance aperture, $a_{in}$, and exit aperture, $a_{out}$,
as well as the length of the concentrator are selected according to the ideal
2-dimensional design proposed in Ref.~\cite{winston}. The cross-section of the
concentrator has an hexagonal shape since this helps to reduce the dead
areas in the camera as opposed to a circular concentrator while at the same time
keeping the same center-to-center distance of pixels.
Figure~\ref{fig:dimensions} shows the dimensions for the entrance aperture and
the output aperture following the geometrical constraints of an open CPC.

\begin{figure}[ht]
\centering
  \includegraphics[ width=.7\textwidth]{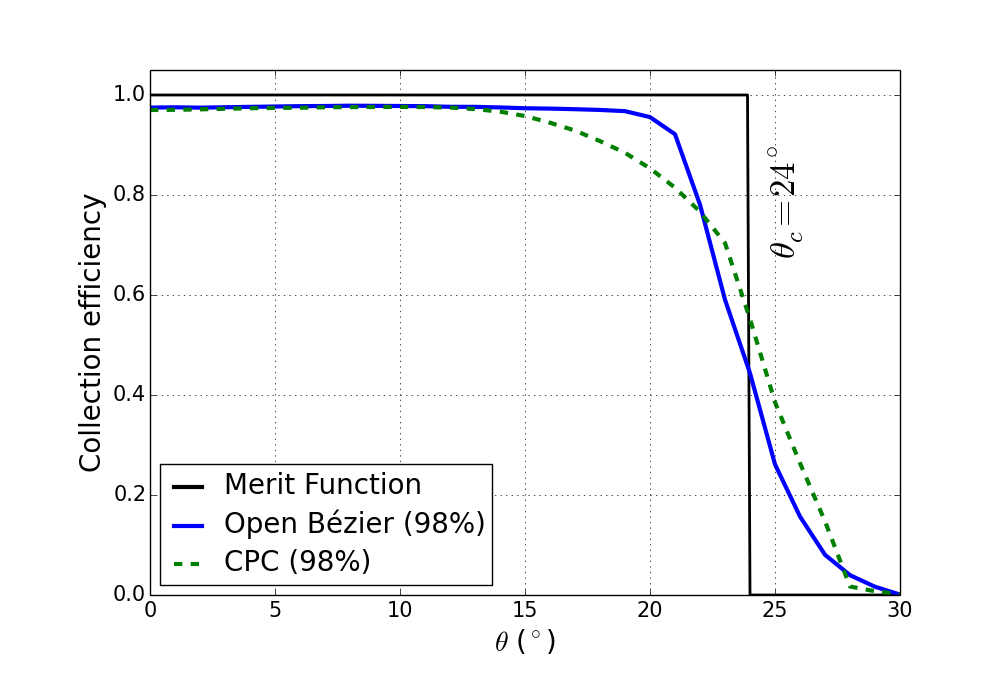}
\caption{Collection efficiency for an open CPC (dotted green) compared to the optimized light concentrator using a cubic B\'ezier curve lateral shape (solid blue). Both simulations have been done with an ideal 98\% reflective material. Also shown is the merit function used in the optimization procedure (black line).}
\label{fig:comparison}
\end{figure}

The sides of the hexagonal concentrator are parametrized using a cubic B\'ezier curve given by:

\begin{equation}
\vec{B}(t) = (1 - t)^3 \vec{P}_{0} + 3(1-t)^2\vec{P}_1 + 3(1-t)t^2\vec{P}_2 + t^3\vec{P}_3, \;\;\; t \in [0,1],
\label{bezier}
\end{equation}

\noindent the B\'ezier curve is described by one parameter, {\it t},
and four control points $\vec{P}_i$. The control points $\vec{P}_0$ and
$\vec{P}_3$ define the output and input aperture sizes respectively, as well
as the length of the collector (see Fig.~\ref{fig:dimensions}) and therefore
they are fixed. In our case geometrical optimization is achieved by moving the control points $\vec{P}_1$ and $\vec{P}_2$. We used the Zemax optimization algorithm based on actively damped least
squares to improve the performance of the optical system. The merit
function of the optimization was set to the collection efficiency of an ideal concentrator (see
Fig.~\ref{fig:comparison}). A similar approach was followed in~\cite{Akira} by
using a brute force optimization by testing several realizations of the light concentrator. 

%JA: I still think the table adds information because gives information about 
%exact shape of the cone, which is not visible in figure 3

\begin{table*}[t!]
\begin{center}
  \caption{Parameters of the B\'ezier function after optimization providing the coordinates of the points shown in Fig.~\ref{fig:dimensions}.}
\vspace{0.2cm}
\small{
\begin{tabular}{cc}
\toprule
Point & Coordinates (mm)\\
\midrule 
$\vec{P}_{0}$  & (5.540, 0) \\
$\vec{P}_{1}$  & (6.434, 2.272) \\
$\vec{P}_{2}$  &  (12.153,  24.048)\\
$\vec{P}_{3}$  &  (13.320, 36.658)\\
\bottomrule
\end{tabular}}
\label{tab:result}
\end{center}
\end{table*}

The results of the optimization are summarized in Tab.~\ref{tab:result}.
Fig.~\ref{fig:comparison} shows the comparison of the collection efficiency with
respect to the standard CPC. In both cases an ideal 98\% reflectivity has been
assumed for the reflective material on the internal surface of the concentrator.
The overall gain in efficiency for angles $\theta < 24^{\circ}$ is about 3\%
while on the other hand the reduction of efficiency for angles $\theta >
24^{\circ}$ is about 30\% which helps to improve the suppression of background
light beyond the cutoff angle.

\begin{figure}[ht]
\centering
  \includegraphics[ width=1\textwidth]{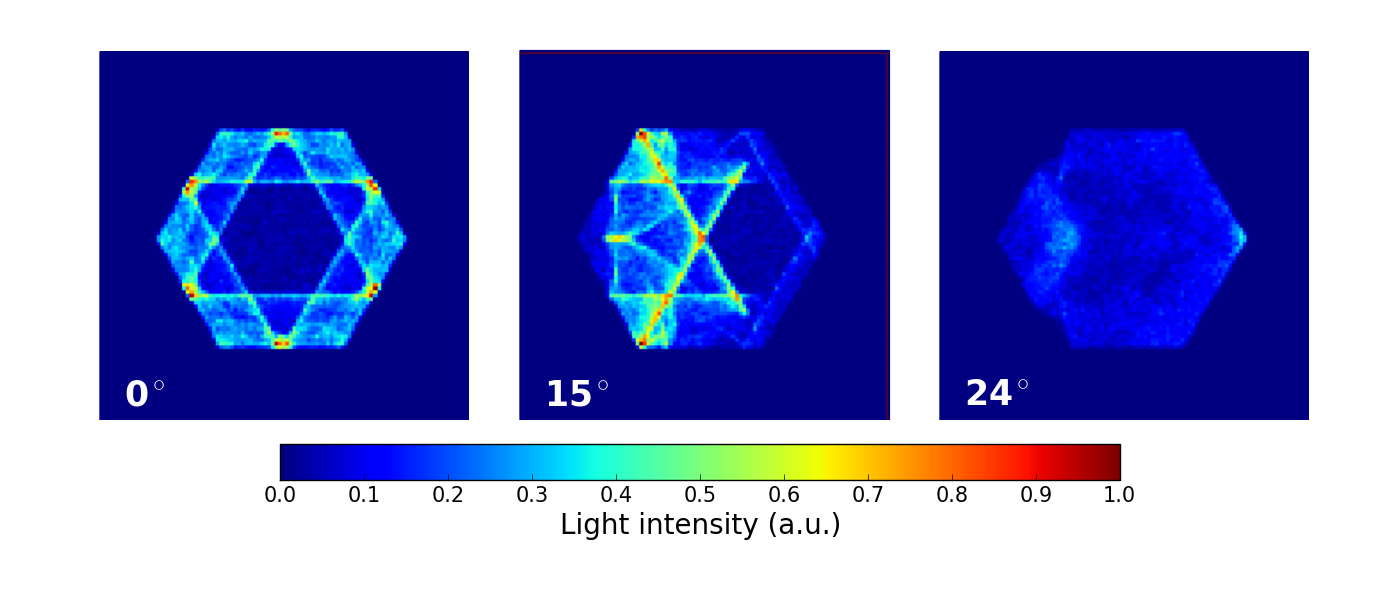}
\caption{Photon distribution at the exit aperture of the light concentrator for
light entering it at three different angles: 0$^{\circ}$ (left),
15$^{\circ}$ (middle) and 24$^{\circ}$ (right).}
\label{fig:photons}
\end{figure}

Both the CPC and the B\'ezier are non-imaging concentrators, meaning that the
distribution of rays at the exit aperture does not reproduce the distribution
of rays entering the concentrator. Fig.~\ref{fig:photons} shows the spatial
distribution of photons at the exit aperture of the B\'ezier optimized
concentrator when a planar waveform of photons enters the concentrator at
three different angles of incidence. The observed non-uniformity will be partly smeared out by the arrival angle distribution of light on the camera plane (see Fig.~\ref{fig:anglescamera}) for different azimuthal angles.

\subsection{Coating simulations}
\label{sec:coating}

An important element of open light concentrators is the reflectivity properties
of the material used to coat the inner surface of the concentrator. In the
previous section we assumed an ideal 98\% reflectivity with no angular or
wavelength dependency. Such an ideal reflector does not exist and therefore more
realistic coatings have been simulated using Zemax until the optimum coating for
this application was selected. 

\begin{figure}[ht]
\centering
\begin{tabular}{c c}
  \includegraphics[ width=.4\textwidth]{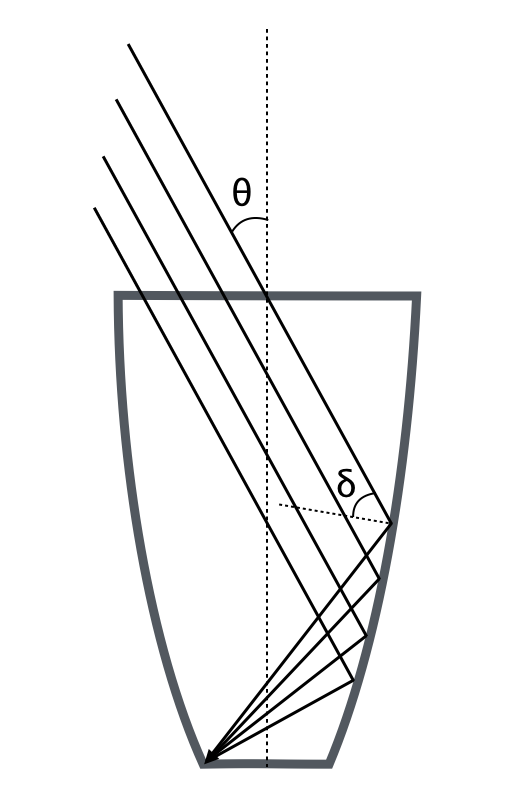}\hspace{0cm}
  \includegraphics[ width=.6\textwidth]{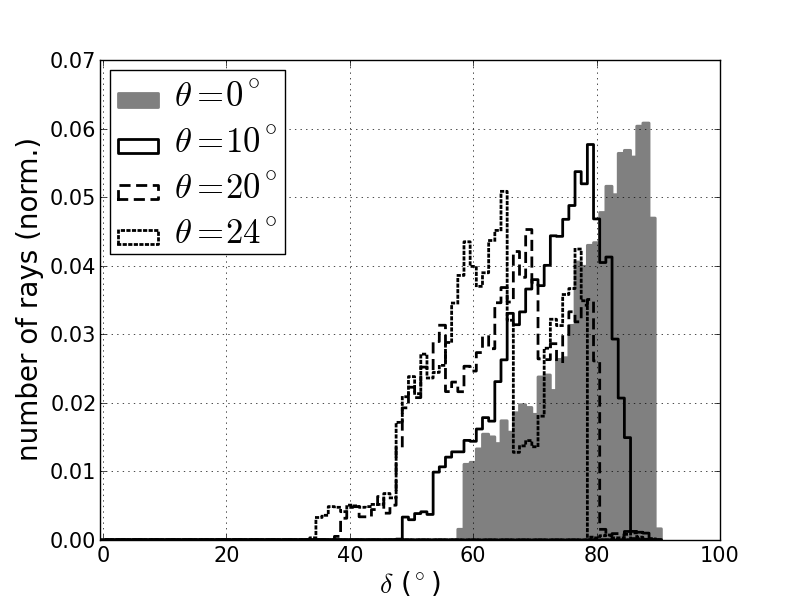}
\end{tabular}
\caption{Left: Scheme showing the path of rays entering the light concentrator in the 2-dimensional case. Right: Distribution of the incident angle on the inner surface of the concentrator, $\delta$, for different entering angles, $\theta$.}
\label{fig:angles}
\end{figure}

The angular dependency of the coating reflectivity has a large impact on the
overall collection efficiency of the concentrator. Using simulations we
were able to estimate the angular distribution of rays on the inner surface of the
concentrator. Figure~\ref{fig:angles} (left) shows a 2-dimensional schematic view of how rays entering into the
concentrator at an angle $\theta$ hit the inner surface at an angle $\delta$. The
right plot on the same Fig.~\ref{fig:angles} shows the distribution of $\delta$ angles in the 3-dimensional case computed from simulations for different incident angles $\theta$. Only rays that 
reach the output aperture are considered. Absorption effects due to
several reflections are not taken into account in this plot. As can be seen,
rays entering perpendicularly ($\theta = 0^\circ$) will impinge on the light
concentrator always at angles $\delta \gtrsim 60^\circ$. For rays arriving with an incident angle equal to the cutoff angle, $\theta = 24^\circ$, all internal reflections occur at angles $\delta
\gtrsim 40^\circ$. Therefore the reflectivity of the coating material for angles
$0^\circ -40^\circ$ is not important for our application. This is an important
information as manufactures of reflective coatings usually seek to have high
reflectivity at normal incident angles, while in our case we aim at high
reflectivity at large angles. 

\begin{figure}[ht]
\centering
  \includegraphics[width=.7\textwidth]{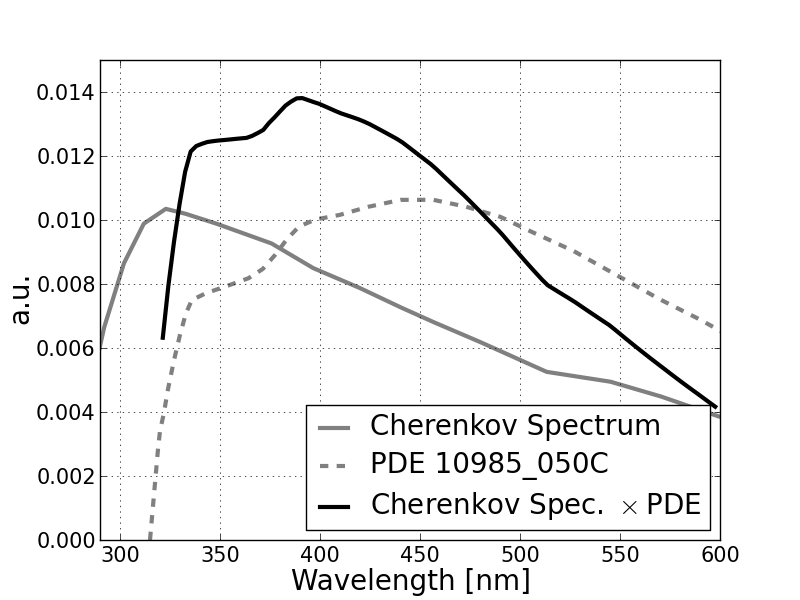}
\caption{The typical Cherenkov spectrum (solid light color line), the PDE of a SiPM (dashed line) and the convolution of the two assumed as light spectrum in Zemax.}
\label{fig:cherenkov}
\end{figure}

The wavelength dependency is also important to optimize the overall efficiency
of the light concentrator. In our application we are interested in the
wavelength range of the Cherenkov spectrum which peaks at the blue and UV light~\cite{CTA} ($< 350$ nm), strongly decreases below 290
nm due to atmospheric transmission effects, and falls off as $1/\lambda^2$
towards longer wavelength (see Fig.~\ref{fig:cherenkov}). The selected
coating needs to have an optimum reflectivity in the wavelength region of the signal convoluted to the response of the photosensor that will be placed at the exit
aperture of the light concentrator. In order to include these two features while
doing the coating optimization using simulation $-$the Cherenkov spectrum and the
detector wavelength response$-$ we simulated a light source with a spectrum which is a
convolution of the two spectra. Since our study is focused on the SST design
using a novel camera equipped with SiPMs solid state sensors (also called Geiger-APD)~\cite{gapds} we used the 
photo-detection efficiency (PDE) of the SiPMs\footnote{The PDE of SiPMs is defined as the product of the Quantum Efficiency
(QE), the fill factor (the ratio of sensitive pixel size over full pixel size, typically of about
60\%) and the avalanche probability (number of excited pixels/number of
pixels hit by photons).} which is wavelength dependent~\cite{gapds}. 

%Again I changed the same form Geiger-APD to SiPM I quote from an email form 
%Adrian:
%naming of G-APD: there seems to by quite some confusion in the
%community. So far, we always called G-APD the full chip of many
%cells. Nevertheless, on the RICH conference it seems most people
%start to use the name 'G-APD' for a single cell, and 'SiPM' for
%the chip consisting of many G-APDs. Probably we should decide
%(and write in the paper(s)) the convention we want to use
%[as far as I remember we did this in the construction paper]

%JA: I think is ok to call it SiPM it will be more confusing to introduce yet 
%another name for SiPM 

In this study we use the PDE of a SiPM from Hamamatsu model
10985\_050C\footnote{www.hamamatsu.org}. The spectrum resulting from the
convolution of the Cherenkov light spectrum and the sensor PDE can be seen in
Fig.~\ref{fig:cherenkov}. This spectrum has a maximum at
wavelengths around 400~nm and therefore the best coating needs to be optimized
for those wavelengths. The NSB on the other hand increases at
higher wavelengths towards the infra-red light~\cite{CTA}. %To further
%suppress the background light above 550 nm, an entrance window coated with a
%wavelength filter will be installed on the camera. 

%JA: we dont mention the windo so this should go on the camera paper: A wavelength
%filter to remove the background region above 550 nm will be applied on the
%entrance window of the camera.

\begin{figure}[ht] 
  \centering 
  \includegraphics[ width=.7\textwidth]{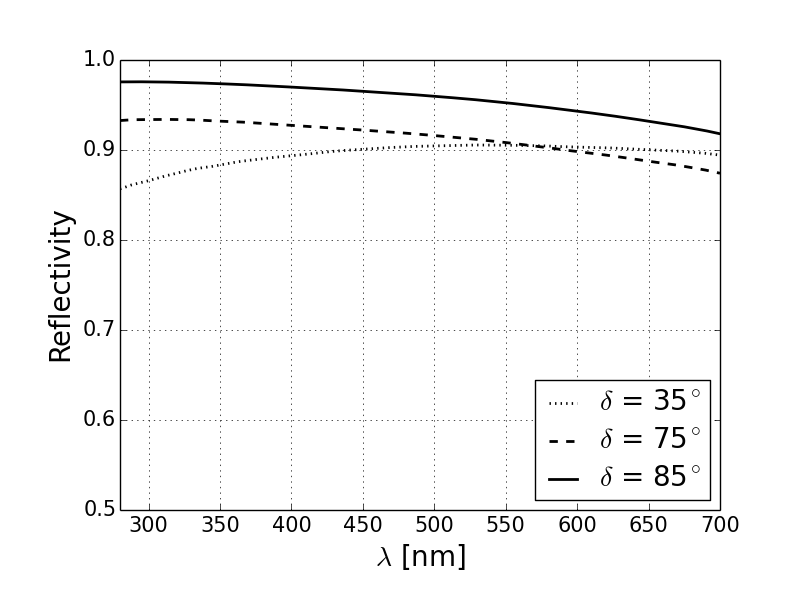}
\caption{Reflectivity as a function of wavelength, $\lambda$, for a R-enhanced
Aluminum coating from Thin Film Physics. Three different incident angles are
shown.} \label{fig:almov8}
\end{figure}

Figure~\ref{fig:almov8} shows the reflectivity of one of the tested materials as
a function of wavelength, $\lambda$ for different incident angles, $\delta$. This coating provided by Thin Film Physics\footnote{THIN FILM PHYSICS (TFP), Regensdorf,
Switzerland.} is made of Aluminum with
superimposed dichroic layers (R-enhanced coating or Al+R) which protect
Aluminum from oxidation and enhance the specular reflection in the region of interest (wavelengths between about 340-550 nm and large incident angles).

\begin{figure}[ht]
\centering
  \includegraphics[width=.7\textwidth]{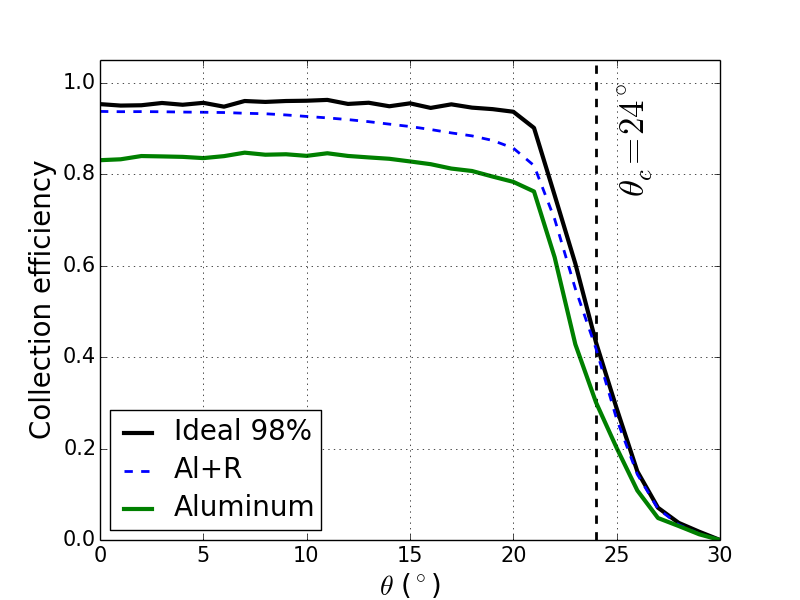}
\caption{Collection efficiency of the light concentrator for different coatings for a light source with spectrum as the one shown in Fig.~\ref{fig:cherenkov}.}
\label{fig:coatings}
\end{figure}

The collection efficiency of the selected coating compared to the ideal reflector of 98\% reflectivity can be seen in Fig.~\ref{fig:coatings}. Also shown is the collection efficiency achievable with a standard
Aluminum coating. The light source used in this comparison emits according to the
spectrum shown in Fig.~\ref{fig:cherenkov}. 
The collection efficiency shown here refers to
the efficiency of the light concentrator standalone.
In this paper we do not discuss the additional effect of an eventual entrance window on the camera since there are multiple possible solutions
that go from thin individual windows on top of single pixels (e.g. 0.7 mm-thick windows) to a common entrance window of the whole camera (e.g. a 3 mm-thick window).
Clearly, depending on the thickness of the material, one expects losses due to absorption (with a thickness dependent cutoff) and due to the Fresnel reflection losses on the contact surfaces between air and the material (which are of the order of 8\%). 
Nonetheless, it should be considered that the reflection losses can be compensated by an anti-reflective coating and that a filter can be applied on the window to cut out the wavelength region
above 550 nm dominated by the NSB. As a matter of fact, in the case of SiPM, the sensitivity stays higher than for PMTs in this optical-IR region.
The combined effect of the light funnels and window will be the subject of a future paper.
Moreover there are additional Fresnel losses also on the silicon layer that protects the photosensor also due to the
different refractive index encountered by light.

\section{Optical measurements of prototypes}
\label{sec:measurements}

A very important element in the design of the light concentrator is the
manufacturing of the substrate since this can severely affect the properties and
reflectivity of the coating. In our design we use the standard
plastic injection molding technique. Being an industrial large scale production
technique, each light concentrator can cost a few tenth of euro centimes. The
requirement in terms of mechanical precision on the shape and the dimension of
the light concentrator are already quite demanding, but the most challenging requirement 
is on the roughness of the inner surface of the concentrator. As
a matter of fact, the coating can just modify/enhance the reflectivity at
different angles and wavelengths, but the overall reflectivity is driven by the
surface smoothness. The desired level of accuracy and roughness was achieved by optimizing the molding injection and polishing technique. %This
%achievement required a certain number of iteration with the German company
%$-$Viaoptics GmbH\footnote{Viaoptics GmbH. Am Leitz-Park 1 | D-35578 Wetzlar,
%Germany}$-$ chosen for its excellent background in plastic injection for optical
%applications.

\begin{figure}
\centering
\includegraphics[width=.7\textwidth]{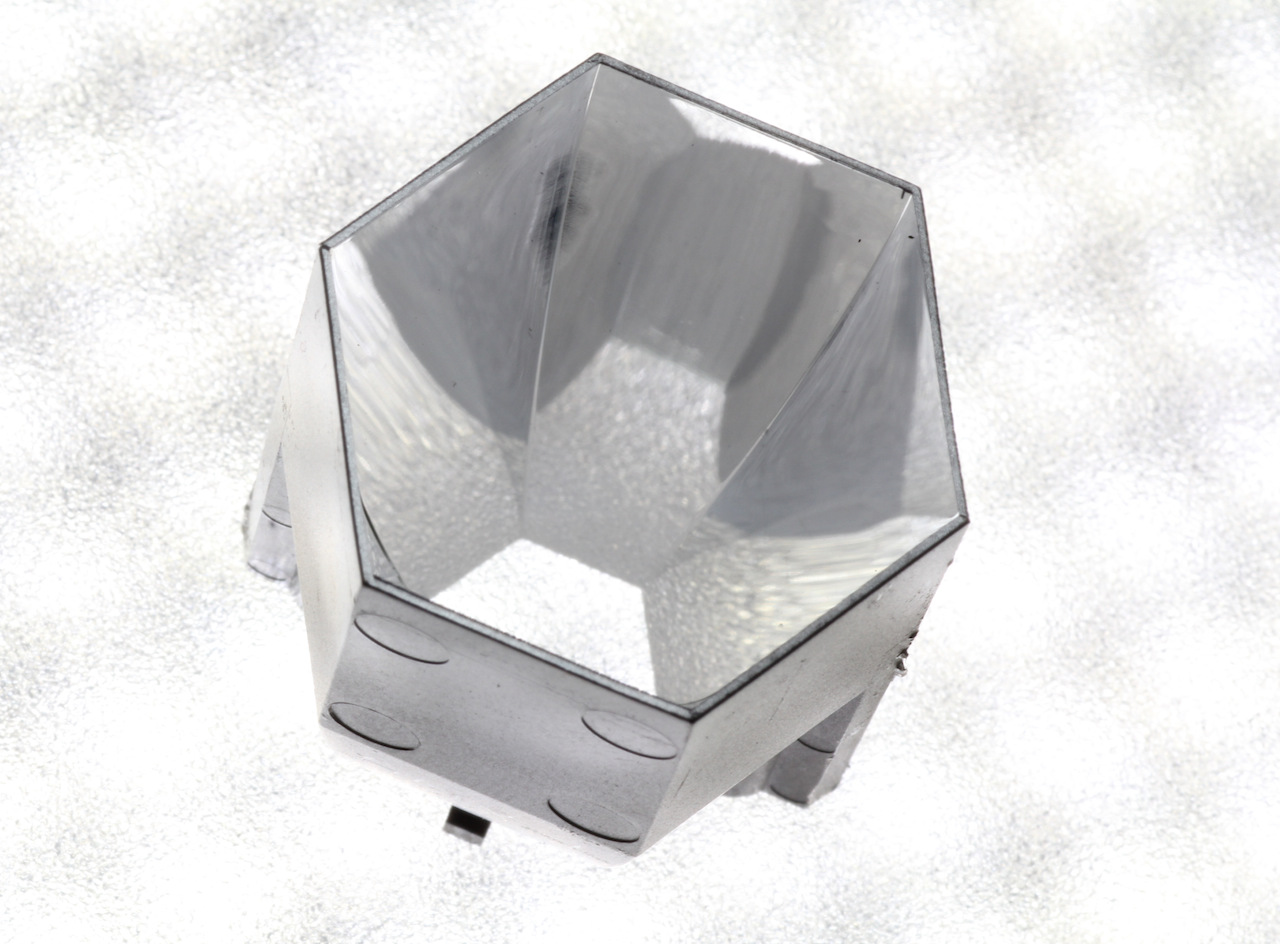}
\caption{Prototype light concentrator manufactured with the injection molding
technique and coated in two halves separately.}
\label{fig:prototype}
\end{figure}

The coating deposition was also
challenging since the main difficulty relies on the fact that the technology
works perfectly on flat surface but it becomes difficult on strongly curved
surfaces as in our light concentrator. In order to facilitate the deposit of the
coating, the light concentrator is manufactured in two halves and then assembled
together after coating. Figure~\ref{fig:prototype} shows a picture of one of the
first prototypes of this technique.

In order to verify that the proposed design achieves the requirements we tested several
prototypes in an optical setup by measuring the absolute collection efficiency
at different wavelengths as a function of the light incident angle. In this section we
will show the measurements corresponding to the optimized concentrator in terms of
geometrical design and coating.

\begin{figure}[ht]
\centering
\begin{tabular}{c}
  \includegraphics[ width=0.8\textwidth]{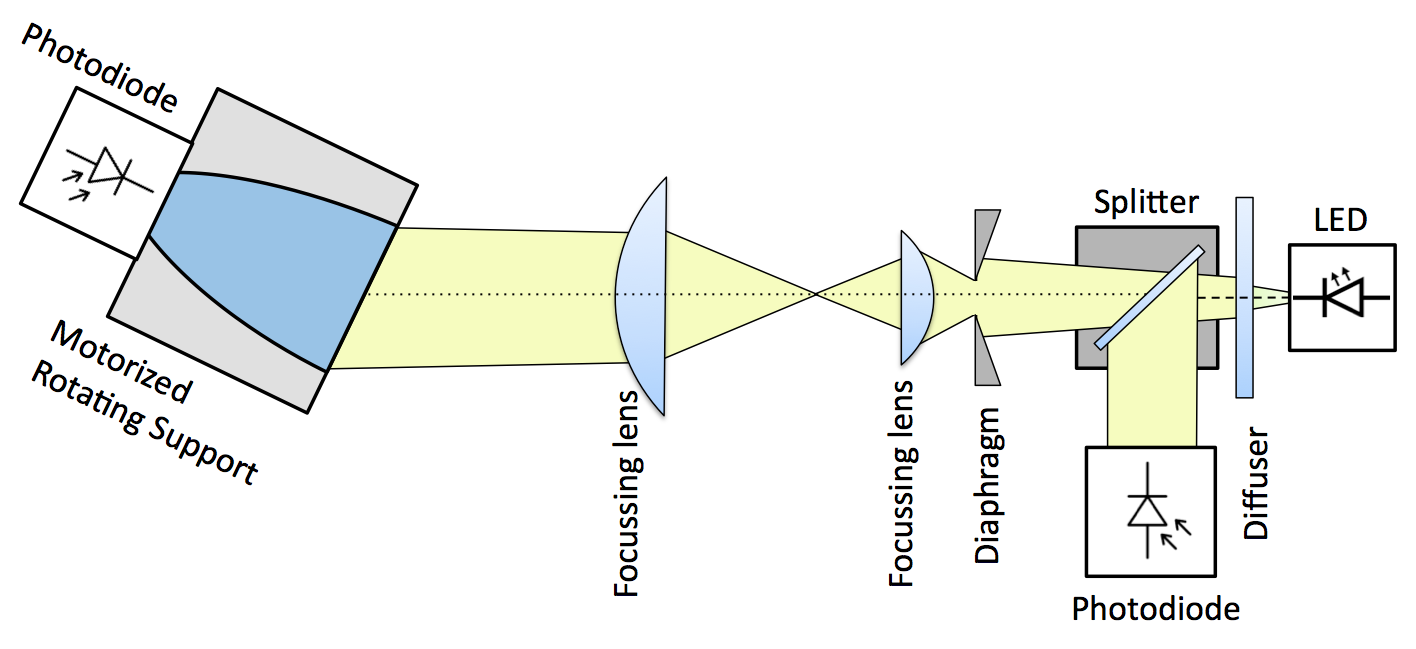}\\
   \includegraphics[width=0.8\textwidth]{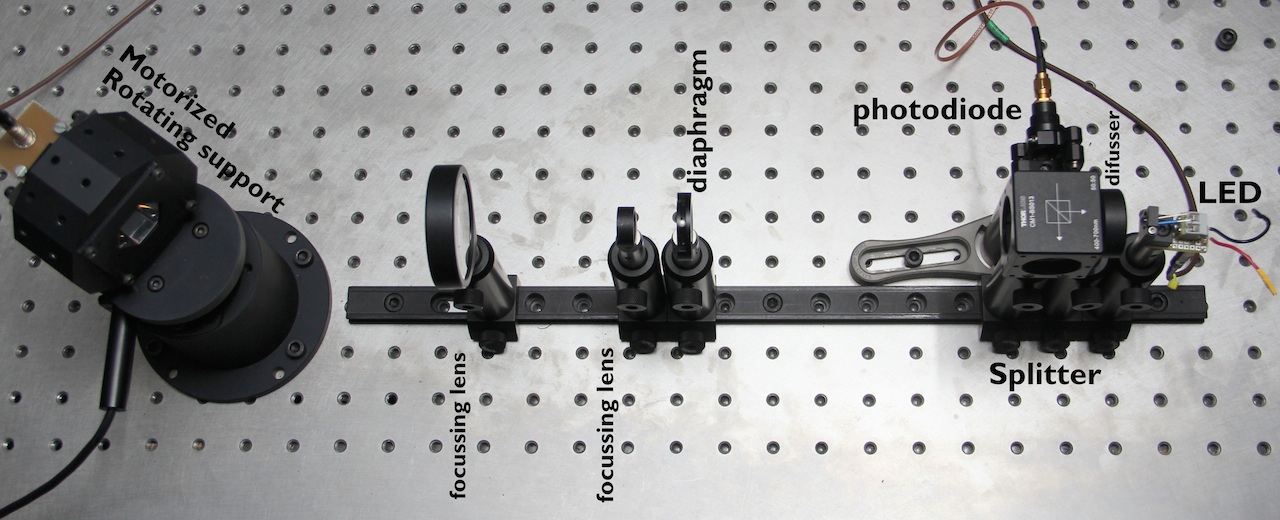}
\end{tabular}
\vspace{0.5cm}
\caption{Optical setup to measure the transmission efficiency as a function of the incident angle. The top figure shows the schematic view of the setup with the different components indicated. The same components can be seen in the photograph of the set-up in the bottom panel.}
\label{fig:optical-setup}
\end{figure}

\begin{figure}[ht]
\centering
  \includegraphics[ width=0.8\textwidth]{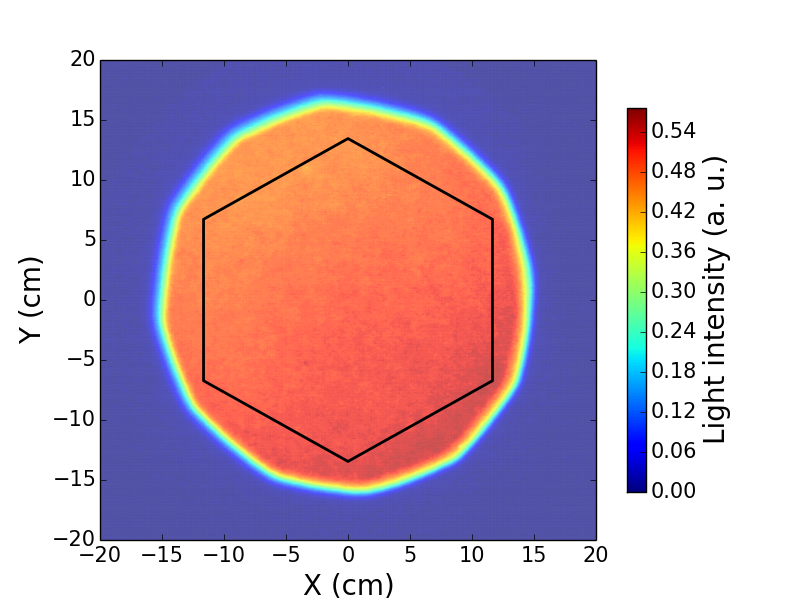}
 \vspace{0.5cm}
\caption{Light intensity of the LED beam at the entrance of the light concentrator. Variations on the uniformity of the light are within  8\%. The hexagon indicates the size of the entrance aperture of the light concentrator.}
\label{fig:beam}
\end{figure}

The experimental set-up for this measurement was
mounted on an optical bench in order to have a perfectly aligned system.  The top panel in Figure~\ref{fig:optical-setup} shows the scheme of the set-up with the
different elements used while the bottom panel shows a picture. The set-up is located in a dark room in order to protect
against ambient light. The light concentrator is placed in a motorized
rotational stage\footnote{Thorlabs: CR1-Z7}. This rotational mount offers
high-precision motion and it is used to study the angular dependency in the
collection efficiency of the light concentrator. The light concentrator is rotated
sequentially from 0$^{\circ}$ to 30$^{\circ}$ while illuminated by an LED. The
light of the LED is collimated by means of different optical elements in order
to have a planar wavefront at the entrance of the concentrator.
Figure~\ref{fig:beam} shows the beam profile as measured at the entrance of the
concentrator. This profile was measured by taking a photograph in RAW mode of the
footprint of the beam on a white surface at the entrance of the light
concentrator using a commercial
Canon 60D.  The dimension of the beam was selected to guarantee the full coverage of the
concentrator entrance aperture. Variations on the uniformity of the light beam inside the region defined by the concentrator aperture are $\sim8$\%.

The light of the LED is recorded by a photodiode used for reference after
splitting the light beam with a 50/50 splitter. This reference diode allows us to
correct from drifts in the light yielded by the LED as a function of time. In
order to calculate the absolute collection efficiency of the concentrator, the
measurement is done in two steps: 

\begin{enumerate}
\item{{\bf Configuration 1.}} In the first step a photodiode is placed on the
           rotational stage at the position of the input aperture of the
           concentrator. The photodiode is then masked with an
           hexagonal shape of the same dimension as the entrance aperture. With this measurement we record the amount of light
           that enters into the concentrator as a function of the incident
           angle.
\item{{\bf Configuration 2.}} In the second configuration the light concentrator
           is placed in the rotation stage and the photodiode is then attached to the
           rear exit aperture of the concentrator measuring the amount of light
           that exits the concentrator as a function of
           the incident angle. \end{enumerate}

\begin{figure}[ht]
\centering 
\begin{tabular}{c c}
  \includegraphics[width=.5\textwidth]{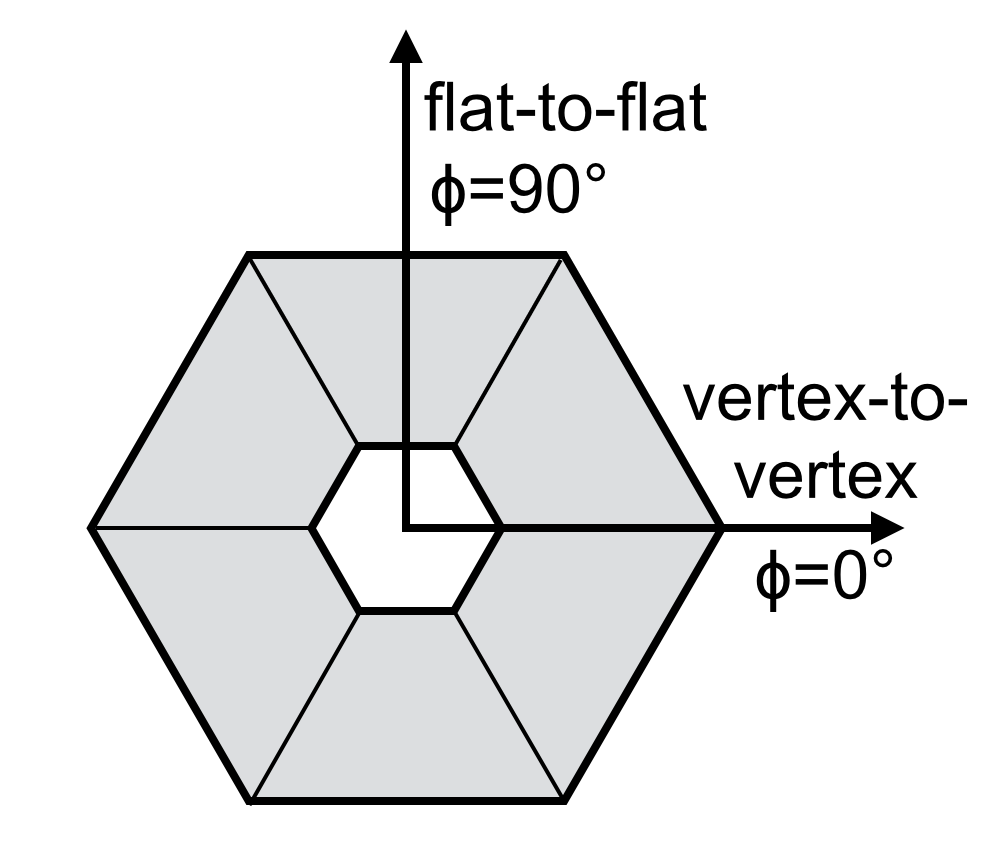}
  \hspace{1cm}
  \includegraphics[width=.3\textwidth]{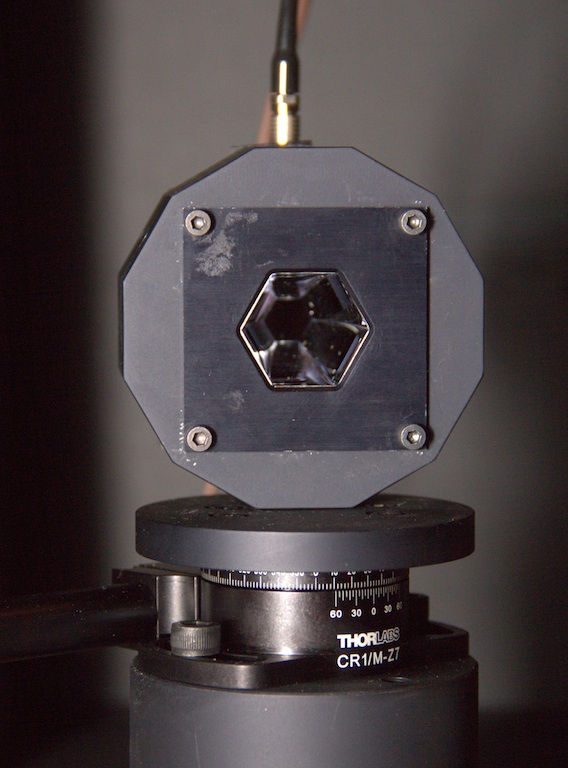}
\end{tabular}
\caption{Left: Scheme of the azimuthal directions in which the light concentrator can rotate. Two main rotation directions are measured, when the concentrator rotates along the plane containing the vertex-to-vertex axis ($\phi = 0^\circ$) and when the concentrator rotates along the plane containing the flat-to-flat axis ($\phi = 90^\circ$). The photograph on the right shows the concentrator mounted on the dodecagon holder. The holder in this picture is in the vertex-to-vertex configuration.}
\label{fig:rotations}
\end{figure}

The absolute collection efficiency is given simply by the ratio of the recorded
light by the photodiode between configuration 1) and 2). Since the entrance
aperture of the light concentrator has an hexagonal shape, there are two major
rotational directions as indicated in Fig.~\ref{fig:rotations} left. In one
direction the light concentrator rotates along the plane containing the vertex-to-vertex axis of
the hexagon, $\phi = 0^\circ$, while the other rotation is along the plane containing the flat-to-flat
axis, $\phi = 90^\circ$. Only rotations along these two major directions are allowed in our set-up.

In order to check any possible coating or geometrical asymmetries in the
azimuthal angle we perform the measurements of the collection efficiency for all
possible major rotation directions: three vertex-to-vertex directions
and three flat-to-flat directions. Figure~\ref{fig:rotations} (right) shows a
picture of the light concentrator in its holder mounted on the rotational
support. The dodecagon shaped holder was manufactured in order to easily change
from one rotational axis to another.

\begin{figure}[h]
\begin{tabular}{c c}
  \includegraphics[ width=0.5\textwidth]{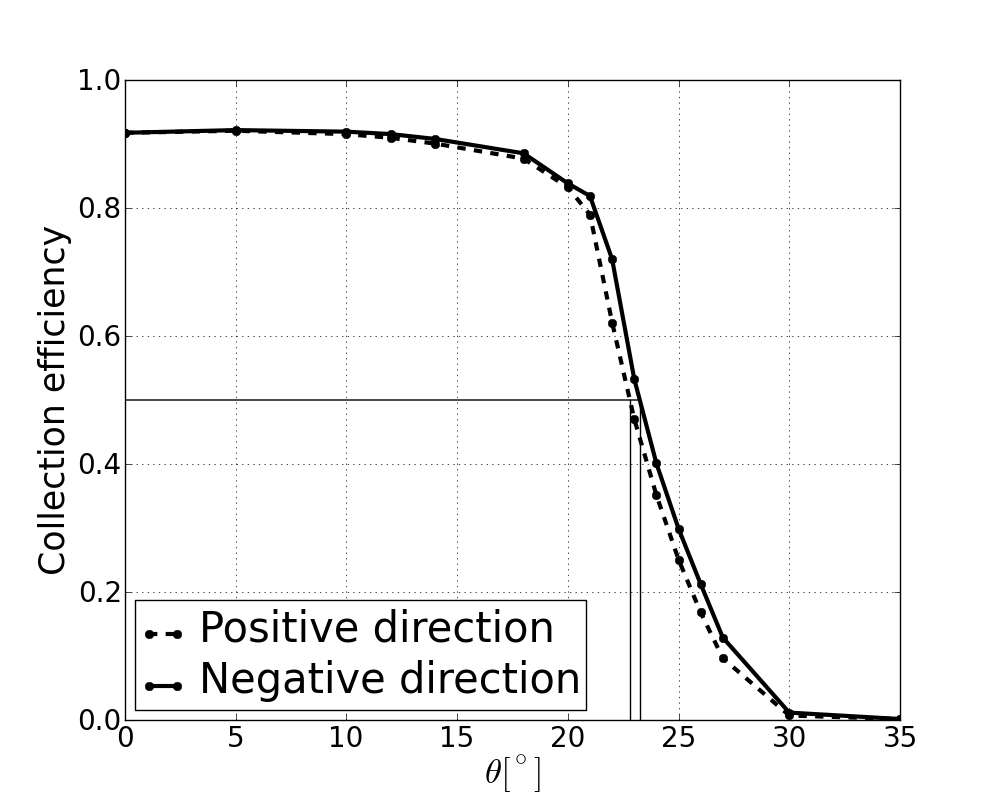}
   \includegraphics[width=0.5\textwidth]{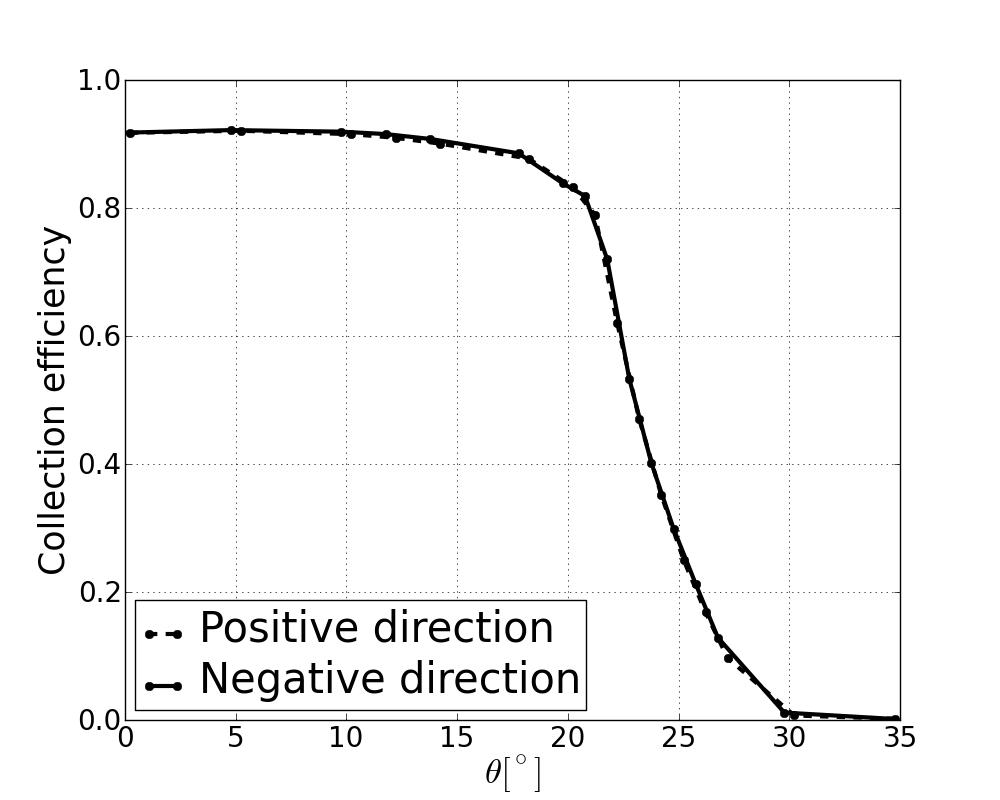}
\vspace{0.5cm}
\end{tabular}
\caption{Collection efficiency for one flat-to-flat measurement for the negative
direction (-35$^{\circ}$ to $0^{\circ}$) and positive direction ($0^{\circ}$ to
$35^{\circ}$) before the angular offset correction (left) and after the offset
correction (right). In our measurements the angular offset applied is typically of the order of $\sim 0.2^{\circ}$.}
\label{fig:alignment}
\end{figure}

As part of the measurement procedure we must correctly define the 0-angle of the
system. In order to align the optical system we first perform an estimation of
the zero by looking at the photon distribution measured by the photodiode as a
function of angle when a pin-hole of 0.5 cm is placed on top of it. This
distribution is strongly peaked and the maximum is used to define the zero of
the system. In addition, each collection efficiency measurement is performed in
an angular range from -35$^{\circ}$ to 35$^{\circ}$, i.e., from the negative to positive angles. This allows us to correct for the zero of the system by simply
applying an angular offset that superimpose the positive-angle values of the collection
efficiency with the negative-angle values. Figure~\ref{fig:alignment} shows this offset correction procedure in one single measurement (flat-to-flat).

\subsection{Results}

\begin{figure}[!h]
\hspace{-1cm}
\begin{tabular}{c c}
\includegraphics[width=0.6\textwidth]{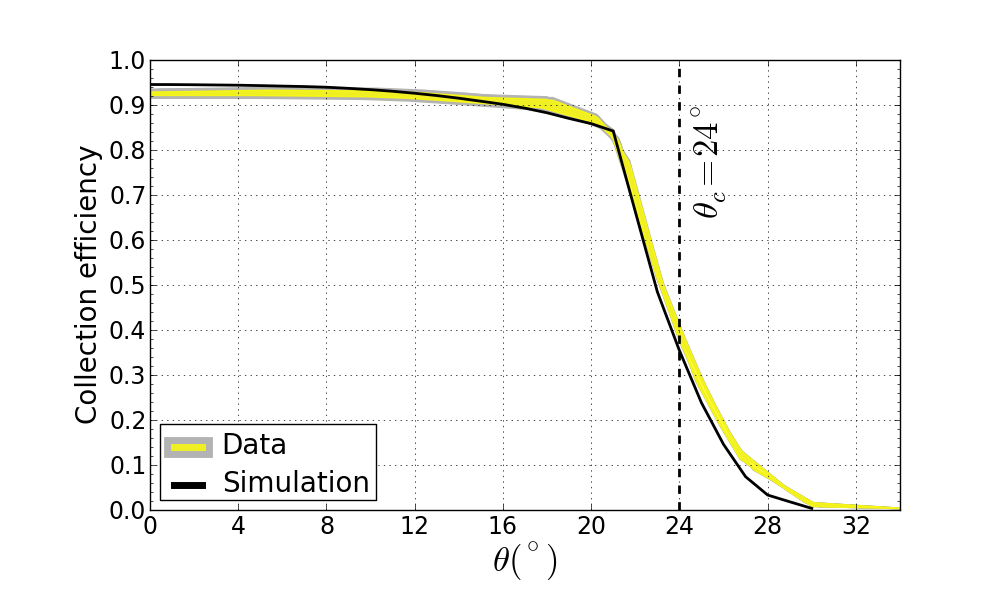}\hspace{-0.7cm}
\includegraphics[width=0.6\textwidth]{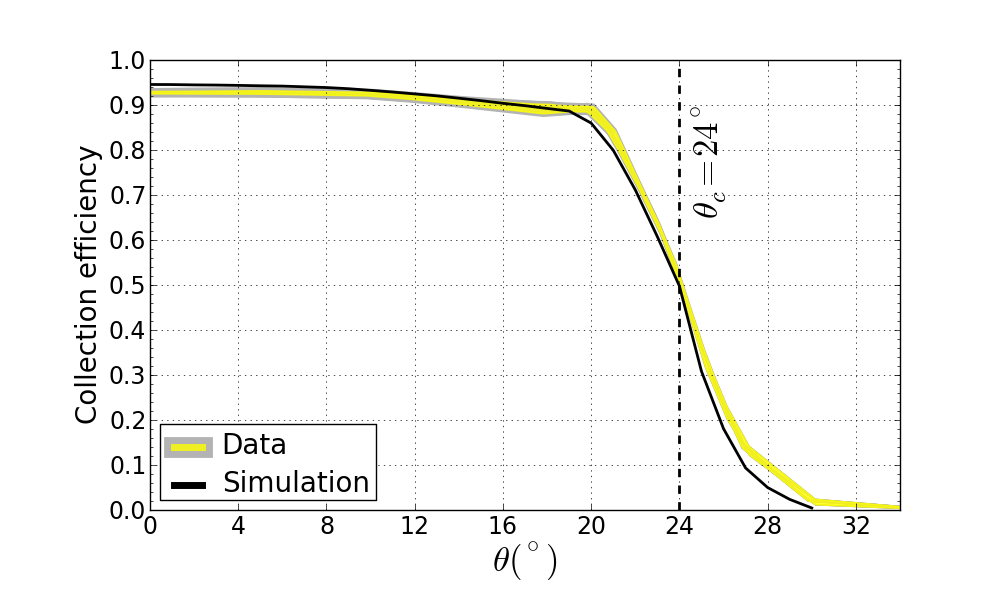}\\
\includegraphics[width=0.6\textwidth]{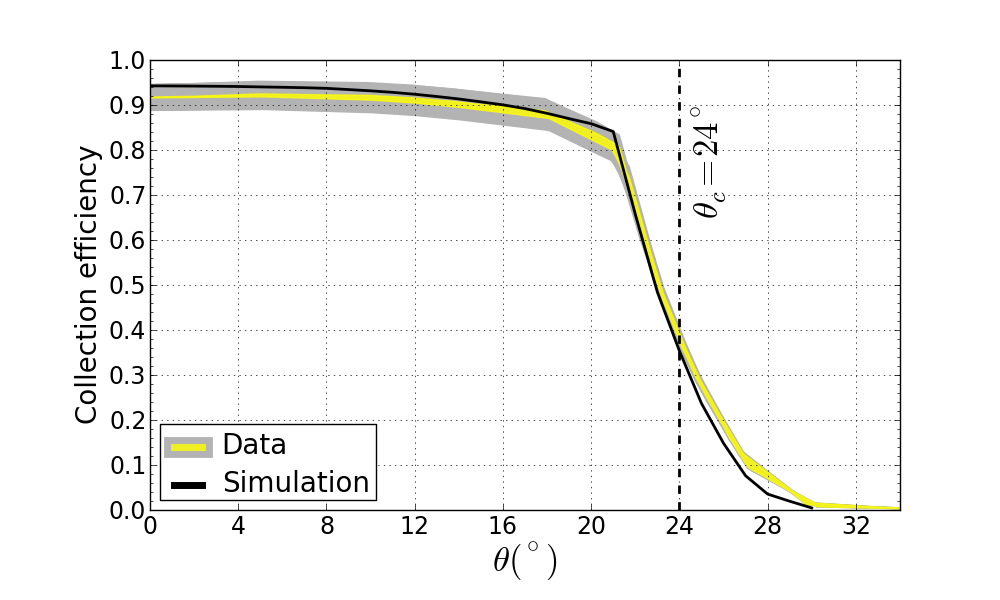}\hspace{-0.7cm}
\includegraphics[width=0.6\textwidth]{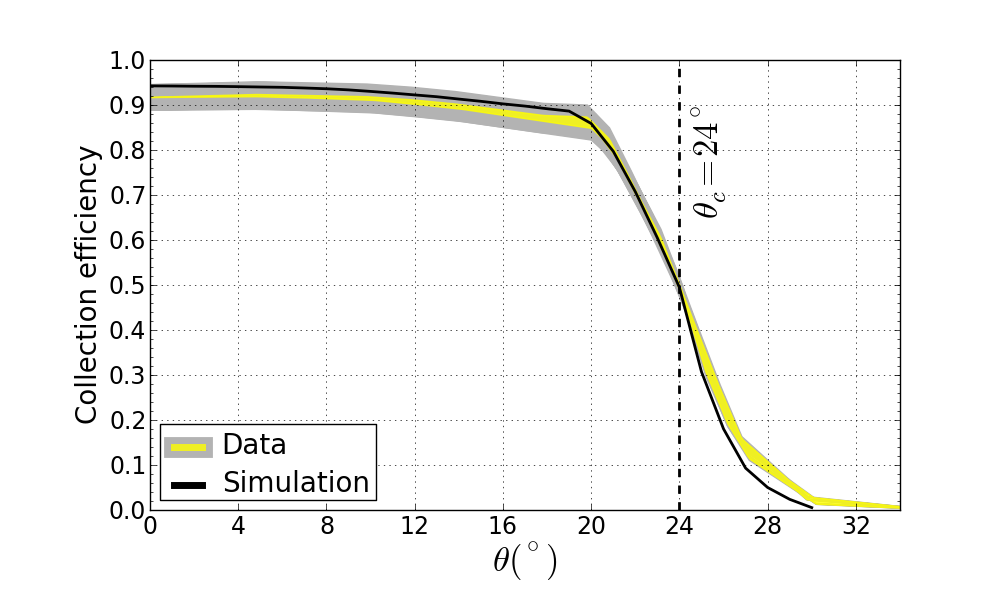}\\
\includegraphics[width=0.6\textwidth]{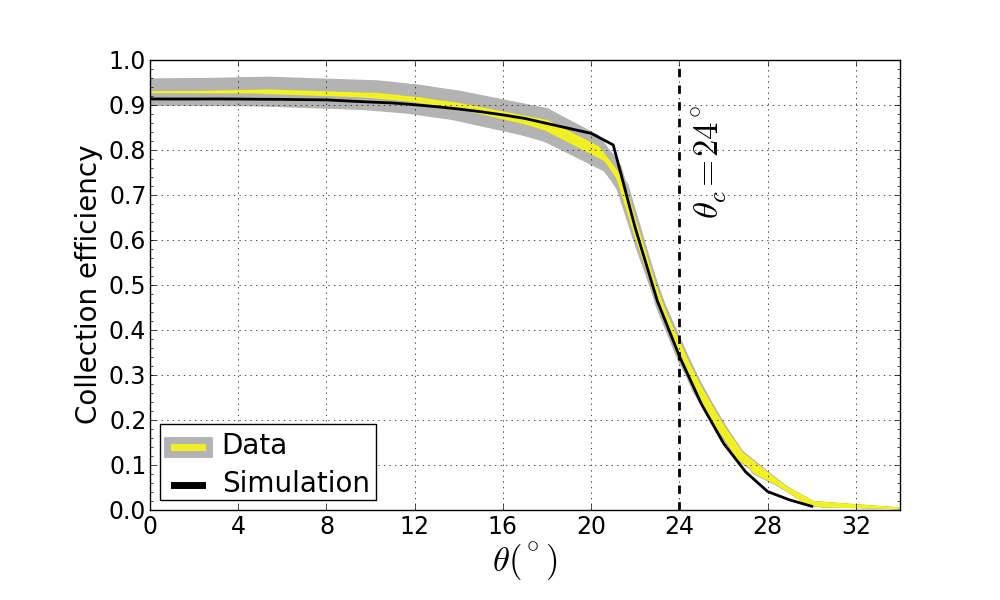}\hspace{-0.7cm}
\includegraphics[width=0.6\textwidth]{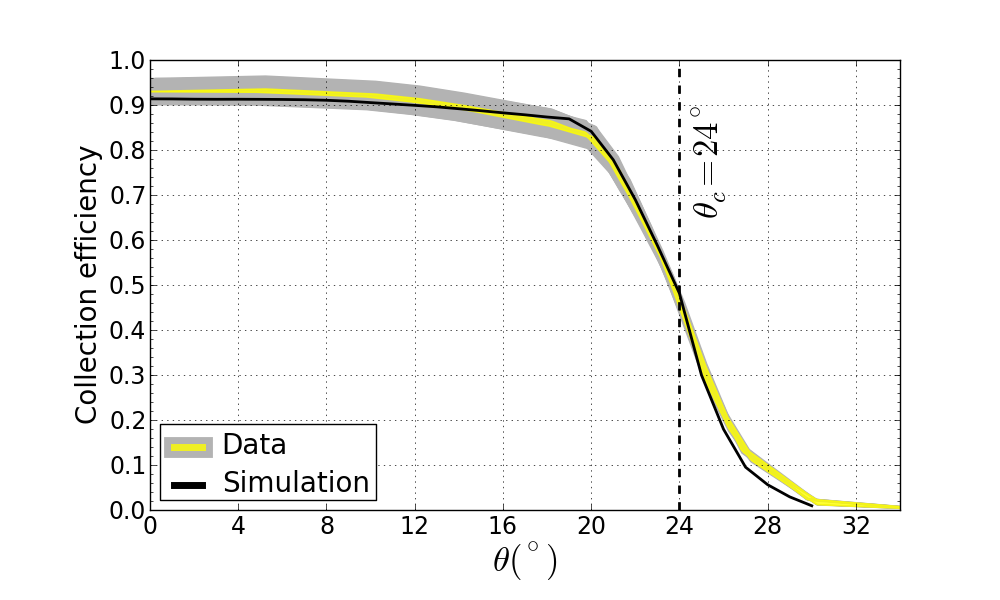}
\end{tabular}
\caption{Collection efficiency measurements for the cone with coating Al+R at a wavelength of 355~nm (top row), 390~nm (middle row) and 595~nm (bottom row). The left column shows the measurements in the flat-to-flat direction and the right plot shows the vertex-to-vertex. The grey area indicates the systematic errors due to the accuracy of the current reading. The black solid line is the corresponding simulation using Zemax for the same wavelength and rotation direction.}
\label{fig:almov8e1}
\end{figure}

Figure~\ref{fig:almov8e1} shows the
collection efficiency for three different wavelengths and two azimuthal rotational
directions (flat-to-flat and vertex-to-vertex). The top row shows the collection
efficiency in the UV with a wavelength of 355~nm, the middle row shows
measurements with a LED of 390~nm and the last row used a LED of 595~nm. The left column
in Fig.~\ref{fig:almov8e1} corresponds to the three measurements in the
flat-to-flat configuration while the right column shows the three
vertex-to-vertex measurements. On each plot the group of three measurements is represented by the
yellow area indicating the range of variability of the measurements. The
picoammeters\footnote{Keithley Picoammeter/Voltage Source 6487} used to measure the generated current from the photodiodes has an intrinsic accuracy that depends on the current level. This
systematic uncertainty is indicated by the grey area. The black solid lines correspond to the simulation using Zemax for the each wavelength and rotation axis. As can be seen the collection efficiency reaches values $>$ 90\% in the
interesting angular region, especially for vertical incident angles ($\theta <
10^{\circ}$).

\begin{figure}
\centering
\includegraphics[width=0.5\textwidth]{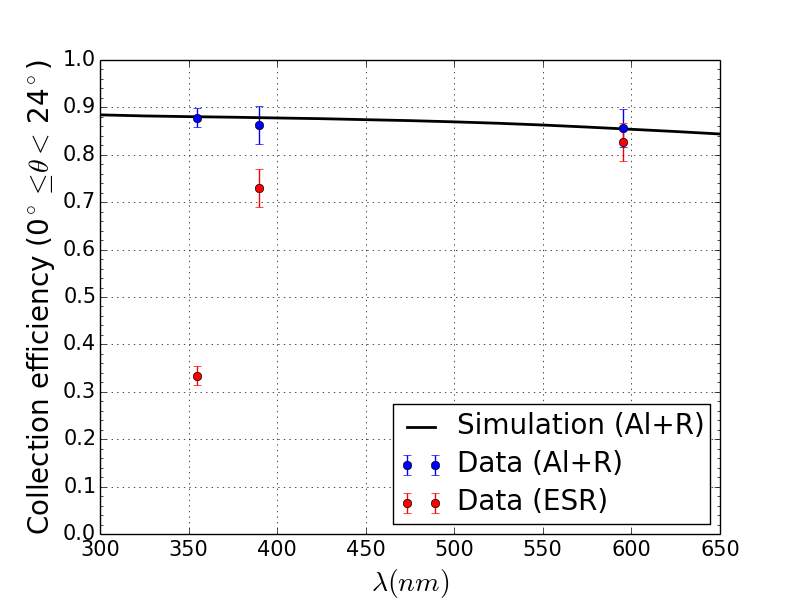}
\caption{Average collection efficiency between 0$^{\circ} \leq \theta < 24^{\circ}$ as a function of the wavelength. Dots are actual measurements while the solid line represents the simulation result. The Al+R coated concentrator is compared with a concentrator with ESR film on it. }
\label{fig:wavelengthdep}
\end{figure}

As pointed out in Sec.~\ref{sec:coating} it is important to have a high
collection efficiency in the UV region since this is where the Cherenkov
spectrum peaks. The measurements shown in Fig.~\ref{fig:almov8e1} show that the
selected coating performs very well in the interesting region of the Cherenkov
spectrum. This is more clearly seen in Fig.~\ref{fig:wavelengthdep}. The
averaged collection efficiency in the angular region of interest (0$^{\circ} - 24^{\circ}$) is shown as a function of wavelength. The black solid
line shows the results from simulation. As a comparison, the
measurements performed on a light concentrator internally covered with the
Vikuiti{\texttrademark} enhanced specular reflector (ESR) film from
3M\footnote{3M SA, Rschlikon, Switzerland.} are also shown. As can be seen the
ESR film performs well outside the UV regime but it has a low
reflectivity in that wavelength range making it not a suitable solution for Cherenkov
astronomy unless it is coated to enhance its reflectivity in the UV. On the contrary, our selected coating has the best collection efficiency in the UV light.

\section{Conclusions}
\label{sec:conclusions}

In this paper we showed the procedure used to design, optimize and manufacture a
cost-effective light concentrator for Cherenkov astronomy.
The selection of coating was driven by simulations and the final measured
collection efficiency achieved a remarkable good agreement with the predictions from simulations. Furthermore, we showed that efficiencies of about 93\% at normal
incident angles are possible with this approach. 

This work was focused on a light guide design suitable for the CTA single mirror small size
telescopes with a camera consisting on SiPM as the active sensors. Nevertheless a
similar optimized design can be adapted for larger size telescopes within the
CTA project independently of the sensors (SiPM or the traditional
photomultipliers) as well as for other projects working in the UV-light seeking to have an increased effective area of the camera such as fluorescence detectors used in cosmic-ray experiments. 

\section*{Acknowledgments}

We acknowledge the Swiss National Science Foundation and the University of
Geneva for the received financial support as well as the Polish grants: 498/FNiTP/158/2011 from the Ministry of Science and Higher Education and
UMO-2011/01/M/ST9/01891 from the National Science Centre for supporting the SST-1M 
project. We also acknowledge Mr. F. Jacocagni from TFP for his
careful feedback and his precious collaboration on the coatings.

%% \appendix{

%% References


\bibliographystyle{elsarticle-num}

\end{document}